\documentclass[conference]{IEEEtran}

\ifCLASSOPTIONcompsoc

  \usepackage[nocompress]{cite}
\else
  \usepackage{cite}
\fi

\usepackage{amsfonts}
\usepackage{wrapfig}
\usepackage[font={small}]{caption}

\ifCLASSINFOpdf
 \usepackage[pdftex]{graphicx}

 \DeclareGraphicsExtensions{.pdf,.jpeg,.png}
\else
   \DeclareGraphicsExtensions{.eps}
\fi

\usepackage{pgfplots}
\usepackage{algorithmic}
\usepackage{algorithm}
\usepackage[fleqn]{amsmath}
\usepackage{amsmath}
\usepackage{tabularx}
\usepackage{multicol, blindtext}
\usepackage{breqn}
\usepackage{multirow}
\usepackage{subfig}
\usepackage{breqn}
\usepackage{color}
\usepackage{graphics}
\usepackage{colortbl}
\usepackage{tabulary}
\usepackage{etoolbox}
\usepackage{colortbl}
\usepackage{bigstrut}
\usepackage{multirow}
\usepackage{array}
\usepackage{tabularx}
\usepackage{booktabs}
\usepackage{hhline}
\usepackage[a4paper]{geometry}
\newgeometry{right=0.62in,left=0.62in,top=0.75in,bottom=1in}

\title{Load Control and Privacy-Preserving Scheme \\ for Data Collection in AMI Networks}

\author{
    \IEEEauthorblockN{Hawzhin Mohammed\IEEEauthorrefmark{1},
	Tolulope A Odetola\IEEEauthorrefmark{1},
	Syed Rafay Hasan\IEEEauthorrefmark{1},
	Mohammad Ashiqur Rahman\IEEEauthorrefmark{7}
	}
    \vspace{0mm}

    \IEEEauthorblockA{\IEEEauthorrefmark{1}Department of Electrical and Computer Engineering, Tennessee Tech University, Cookeville, TN, USA}
    \IEEEauthorblockA{\IEEEauthorrefmark{7}Department of Computer Science, Tennessee Tech University, Cookeville, TN, USA}

\vspace{-7mm}
}

\begin{document}

\maketitle

\begin{abstract}

In Advanced Metering Infrastructure (AMI) systems, smart meters (SM) send fine-grained power consumption information to the utility company, yet this power consumption information can uncover sensitive information about the consumers' lifestyle.
To allow the utility company to gather the power consumption information while safeguarding the consumers' privacy, different methods that broadly utilize symmetric key and asymmetric key cryptography operation have been generally utilized.
%Notwithstanding, these strategies normally include expansive overhead as far as computation and communication.
In this paper, we propose an effective method that uses symmetric key cryptography and hashing operation to gather power consumption information.
Moreover, provide the utility company with an overview of the type of the appliances used by its power consumer and range of power use.
The idea is based on sending cover power consumption information from the smart meters and removes these covers by including every one of the smart meters' messages, with the goal that the utility can take in the accumulated power consumption information, yet cannot take in the individual readings.
%We additionally present a key management methodology that utilizations unbalanced key activities, however, not at all like the power utilization accumulation that is done exceptionally often, the key administration method is run each lengthy timespan for key recharges.
Our assessments show that the cryptographic operations required in our scheme are substantially more effective than the operations required in other schemes.
%In addition, we have demonstrated that the proposed plan can safeguard the consumers' security and give high assurance level against intrigue assaults.

\end{abstract}


%%% ----------------------- Introduction ----------------------- %%%
\vspace{0mm}
\section{Introduction}
\label{Introduction}
\vspace{0mm}

The smart grid utilizes smart devices that have communication and also calculation capacities to update the current power grid.
This new update intends to utilize power more productively, diminish the CO$_2$ emissions, and integrate sustainable power source assets.
One of the important applications in the smart grid that is as of now being actualized is the Advanced Metering Infrastructure (AMI) \cite{gungor2013survey}.

The AMI systems have smart meters placed on the customer side to associate the consumers with the service provider (utility company).
The smart meters send fine-grained power consumption information to the utility company to screen the power requests over brief periods.
The frequent of the information transmission can be as little as a couple of minutes.
However, the fine-grained power consumption information being gathered can be utilized to explain the exercises and behavior of the consumer \cite{garcia2010privacy, alanwar2017proloc}.
It can uncover the kinds of exercises going ahead in the house like the appliances that are right now utilized, regardless of whether a consumer is voyaging, when consumer out of the home and when they return back, and so on.
Clearly, uncovering such sensitive information is a risk to the privacy protection of the consumers, in the long run, this privacy problem may make them quit from demand response and AMI systems \cite{gong2016privacy, li2014eppdr}.

To address this privacy issue, a few ideas have been proposed as of late to empower the utility company to gather the fine-grained power consumption information while protecting the consumers' privacy and security \cite{bos2017privacy, li2010secure, lu2012eppa, saputro2014preserving}.
The vast majority of these schemes either widely utilize asymmetric key cryptography operation, for example, homomorphic encryption strategies or depend on control batteries to conceal the real power utilization.
Nevertheless, these strategies are expensive as far as equipment or communication and computation.
Specifically, due to utilizing symmetric key cryptography in the gathering of readings, ciphertext size is vast and calculation time is long in light of the fact that they require costly calculations.

In this paper, we propose an effective method that for the most part utilizes lightweight symmetric key cryptography and hashing operations to gather power consumption information while safeguarding consumer's privacy.
The thought depends on sending conceal power consumption information in a vector manner by the smart meters and expelling these covers by aggregating every one of the smart meters' messages, with the goal that the utility company can take in the accumulated reading, however, the utility company cannot learn the individual power consumption information.
Each smart meter shares secret cover values with the utility company.
The smart meter covers its power consumption information with the shared secret cover values with the utility company.
Keeping in mind the end goal is to remove the covers by the utility company and extracting the accumulated information form every one of the smart meters.
Every intermediary smart meter adds to its power consumption information the shared secret cover value with the utility company.
To get the aggregated power consumption information by the utility company, all the shared secret cover values shared with all smart meters must be subtracted from the covered aggregated power consumption information received by the utility company.
So, to process the collected power consumption information without knowing the smart meters' individual power consumption information, all the smart meters information should be added.

In our scheme, each smart meter should share a symmetric key with the utility company to secure the communications.
To allow the smart meter and utility company to process an alternate secret cover value for each power consumption information reporting, they should share a symmetric key and utilize it to calculate the shared secret cover value.
Proficient hashing operations can be utilized to calculate the shared secret cover value.
We additionally present a key management system that utilizations asymmetric key operation, yet dissimilar to the power consumption information gathering that is done oftentimes, the key management system is run each prolonged range of time for key reestablishments.

Our calculations show that the cryptographic operations required in our scheme are substantially more effective than the operation required in the present schemes.
Our investigation demonstrates that the proposed scheme can safeguard the consumers' privacy and give high-security level against collusion attacks.
Moreover, there is an intriguing tradeoff between privacy level protection and system overhead.
Utilizing more privacy level protection increments the overhead in size of the shared keys and covers, however, it can give more insurance against attackers.
We have additionally directed experimentation on MatLab to evaluate our scheme and calculate the overhead of the system.

The remainder of this paper is organized as follows.
The Network and Adversary Models are described in Section \ref{Network and Adversary Models}.
The Load Control and Privacy-Preserving Scheme explained is in Section \ref{Proposed Scheme}.
The results are explained in Section \ref{Results}.
Conclusions is drawn in Section \ref{Conclusion}.
%%% =========== Introduction =========== %%%

%%% ----------------------- Figure ----------------------- %%%
\begin{figure}[t]
\setlength{\abovecaptionskip}{0mm}   % 0.5cm as an example
\setlength{\belowcaptionskip}{0mm}   % 0.5cm as an example
\centering
\includegraphics[scale=0.29]{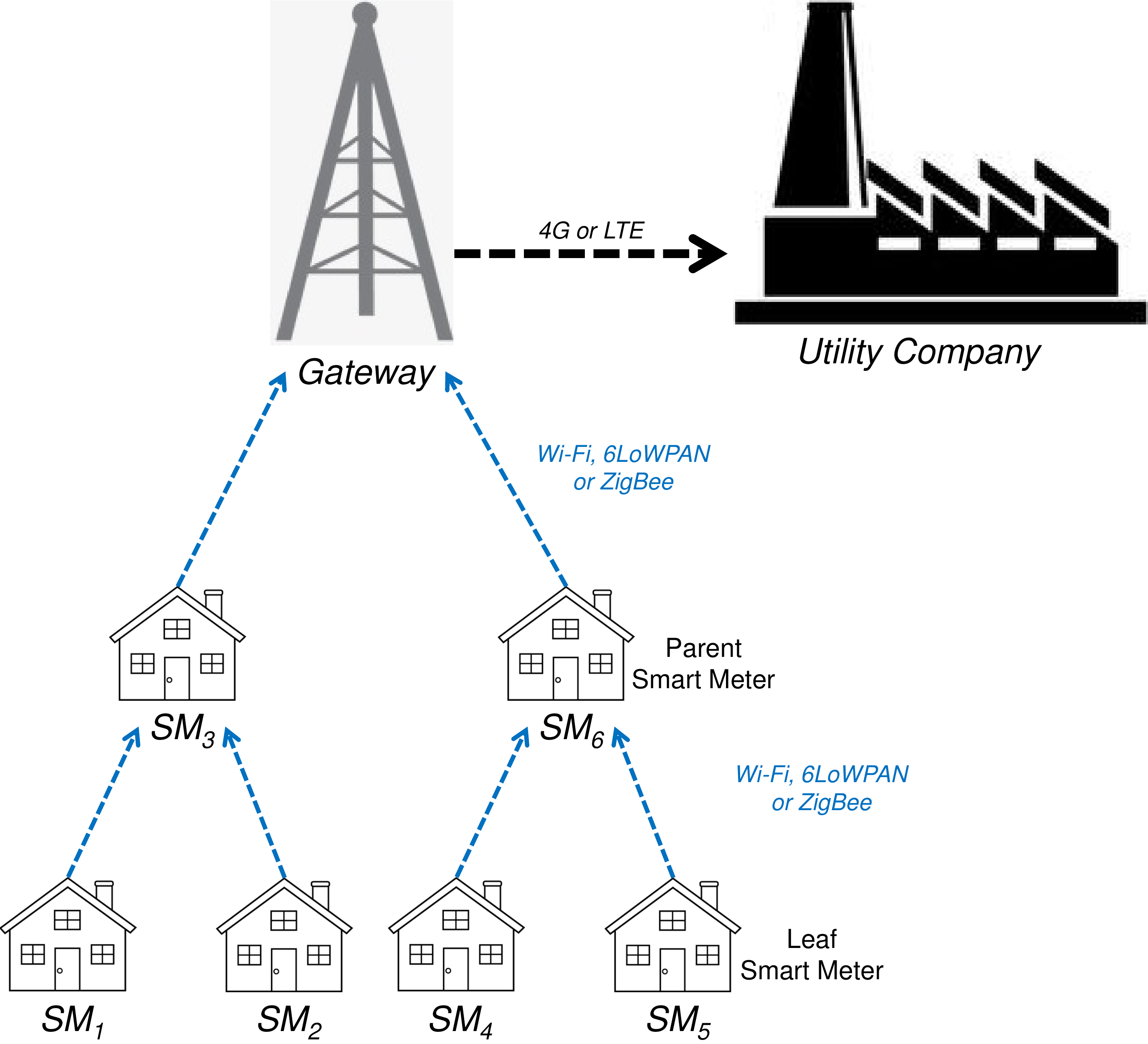}
\caption{Tree topology network in (Multi-hop) AMI network.}
\vspace{-5mm}
\label{fig:Multi-hop}
\end{figure}
%%% =========== Figure =========== %%%

%%% ----------------------- NETWORK AND ADVERSARY MODELS ----------------------- %%%
\vspace{0mm}
\section{Network and Adversary Models}
\label{Network and Adversary Models}
\vspace{0mm}

In this section, we describe the network and adversary model.
We explain how each entity is connected in the network model and what is the aim of the attacker in the adversary model.

\vspace{0mm}
\subsection{Network Model}
\label{Network Model}
\vspace{0mm}

The system network model consists of smart meters (SMs), Gateway, and utility company \cite{gungor2013survey, gungor2011smart}.
The SMs are linked to each other and to the Gateway utilizing short-range connection protocol like Wi-Fi, ZigBee or 6LoWPAN, as shown in Fig. \ref{fig:Multi-hop}.
Smart meters cannot be linked straight to the Utility company, however, the Gateway falls in as a hand-off between the smart meters and the Utility company.
Each smart meter operates as a switch to hand-off smart meters' packets to link them to the Gateway.
The Gateway gathers all smart meters power consumption information and sends an accumulated power consumption information to the Utility company through a long-range connection, for example, 4G, LTE, or fiber optic links.
This power consumption information transmission rate can be less than a minute for a commercial building and more than a minute for household building \cite{zheng2013smart}.

%%% ----------------------- Figure ----------------------- %%%
\begin{figure}[t]
\setlength{\abovecaptionskip}{0mm}   % 0.5cm as an example
\setlength{\belowcaptionskip}{0mm}   % 0.5cm as an example
\centering
\includegraphics[scale=0.29]{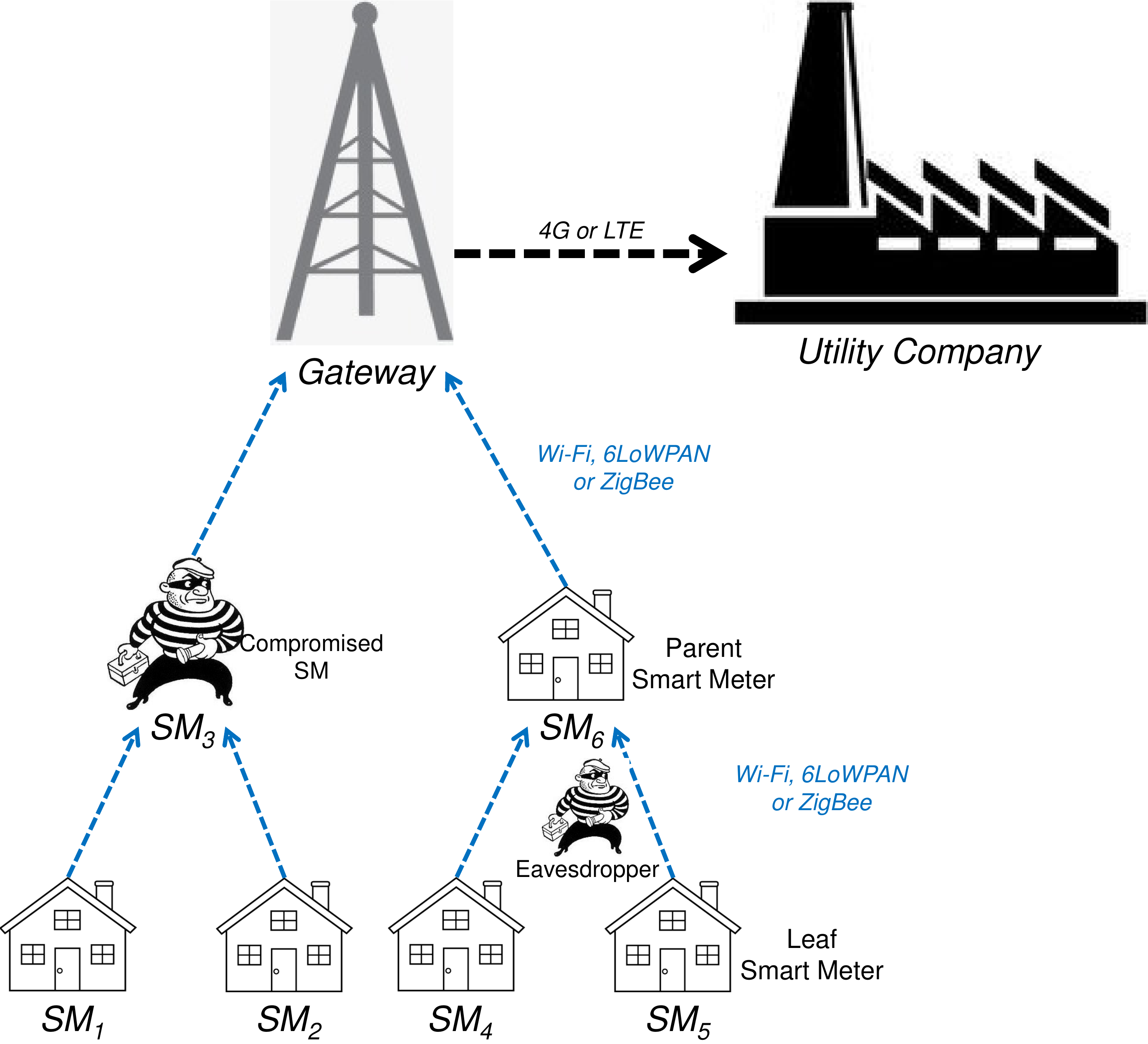}
\caption{Adversary Model, Compromised SM and Eavesdropper.}
\vspace{-5mm}
\label{fig:AdversaryModel}
\end{figure}
%%% =========== Figure =========== %%%

\vspace{0mm}
\subsection{Adversary Model}
\label{Adversary Model}
\vspace{0mm}

We examine fair-but-curious adversary model.
The attacker is interested in learning private data on the power consumers yet the attacker would prefer not to bother the communication or the appropriate operation of the system network.
We consider a strong attacker model by expecting that attacker can be both interior and/or exterior and can be all smart meters, as shown in Fig. \ref{fig:AdversaryModel}.
The exterior attacker can observe the transmissions messages and the interior attacker can compromise other smart meters to take in the secret cover values they include for covering their power consumption information.
For cooperation attack(i. e., collusion attack), we expect that any number of interior and/or exterior compromised smart meters can cooperate, however, this cooperation can be at most \textit{n-1} smart meters, where the Utility company has \textit{n} smart meters.
The objective of the attacker is to figure out how many smart home appliances are there in the community, how much power they are consuming from the grid and in which level they are.

%%% =========== NETWORK AND ADVERSARY MODELS =========== %%%

%%% ----------------------- Figure ----------------------- %%%
\begin{figure}[t]
\setlength{\abovecaptionskip}{0mm}   % 0.5cm as an example
\setlength{\belowcaptionskip}{0mm}   % 0.5cm as an example
\centering
\includegraphics[scale=0.27]{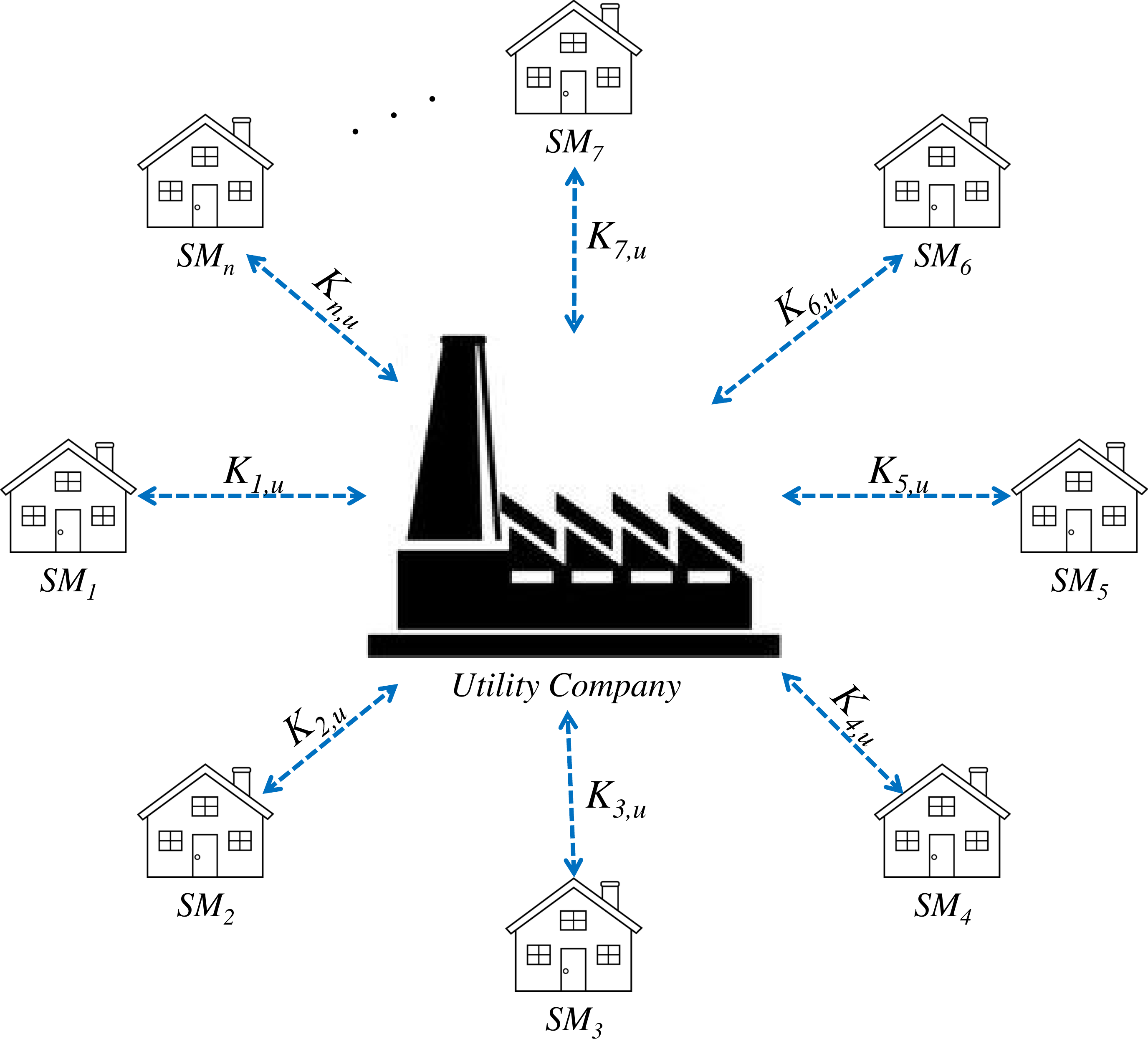}
\caption{Key Agreement Topology, Each SM share a Symmetric Key with the Utility Company.}
\vspace{-5mm}
\label{fig:KeyAgreement}
\end{figure}
%%% =========== Figure =========== %%%

%%% ----------------------- PROPOSED SCHEME ----------------------- %%%
\vspace{0mm}
\section{Load Control and Privacy-Preserving Scheme}
\label{Proposed Scheme}
\vspace{0mm}

The scheme is described in this section.
We start with an overview of the scheme, then the system setup, after that we explain how the key management is working followed by the power consumption vector generation and finally the load control by the utility company.

\vspace{0mm}
\subsection{Overview}
\label{Overview}
\vspace{0mm}

Each smart meter $SM_i$ shares a symmetric key with the Utility company.
These keys are utilized to calculate the secret cover value to cover the power consumption information send by the smart meters, as shown in Fig. \ref{fig:KeyAgreement}.
Each smart meter covers its power consumption information for the privacy issue, and the secret cover value subtracted by the Utility company from the accumulated power consumption information to get the total power consumption information reading without knowing the individual readings.

In this section, we initially disclose the required cryptography to bootstrap our scheme system.
Then, we interpret the way of including/expelling secret cover value and explain the information accumulation scheme in our systems.
At last, we interpret a productive and secure key management strategy.

\vspace{0mm}
\subsection{System Setup}
\label{System Setup}
\vspace{0mm}

A trusted authority (TA) generates a finite field $Z^*_ q$ of order $q$, where $q$ is a large prime number.
$G_1$ is additive group and $P$ is the generator of the group $G_1$.
$E_k()$ and $D_k()$ are symmetric-key encryption and decryption operations, where $k$ is the secret shared key.

There is one one-way hash functions $H_1$, where $H_1: {\{0, 1\}}^* \rightarrow$ $G_1$.
%, $H_2: {\{0, 1\}}^* \rightarrow  Z^*_q$ and a keyed hash function $H_K : {\{0, 1\}}^* \rightarrow {\{0, 1\}^t}$, where $K$ is a secret key used to calculate the hash value and $t$ is the $HMAC_K$’s output size.
%We utilize the keyed hash function HMAC in our system because it is effective and generally adopted.
For each smart meter $SM_i$, the TA picks a random element $sk_i \in Z^*_q$ and calculates $PK_i = sk_iP$, where $sk_i$ and $PK_i$ are the smart meter’s secret and public keys, respectively.

Paillier homomorphic cryptosystem \cite{paillier1999public, bae2016preserving} also has been utilized in our scheme for power consumption accumulation in a privacy-preserving way.
Homomorphic encryption is a sort of public key encryption in which multiplying two encrypted messages (while in cypher domain) results in the summation of the messages (in plaintext domain).
The Utility company produces homomorphic public key ($N$, $G$) and private key ($\lambda$)
\begin{equation*}
\textit{Homomrphic encryption} = G^m \zeta^N  \textit{mod } N^2
\end{equation*}
The TA distributes the public parameters of the system $\{q, P, H_1,  E_k(), D_k(), N, \textit{and } G\}$.

%%% ----------------------- Figure ----------------------- %%%
\begin{figure}[t]
\setlength{\abovecaptionskip}{0mm}   % 0.5cm as an example
\setlength{\belowcaptionskip}{0mm}   % 0.5cm as an example
\centering
\includegraphics[scale=0.27]{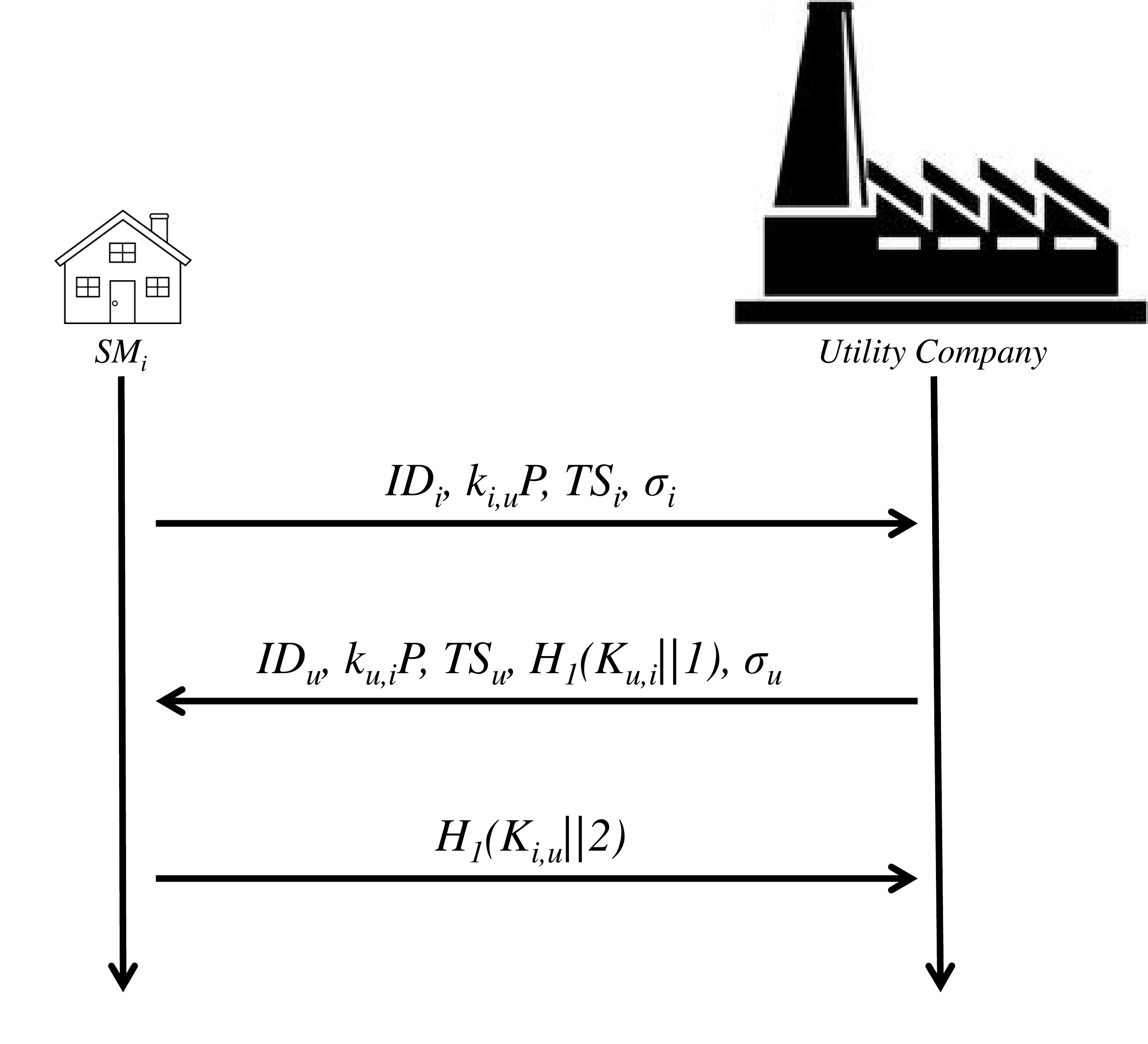}
\caption{Key Agreement steps between SM and the Utility Company.}
\vspace{-5mm}
\label{fig:KeyAgreement_1}
\end{figure}
%%% =========== Figure =========== %%%

\vspace{0mm}
\subsection{Key agreement procedure}
\label{Key agreement procedure}
\vspace{0mm}

The key agreement procedure should be executed in one of the following cases:
\begin{itemize}
\item At the bootstrap of the system;
\item When a new smart meter is added/removed to/from the AMI networks;
\item In key renewals;
\end{itemize}
Each smart meter $SM_i$ needs to share a symmetric key with the Utility company.
As explained earlier, the symmetric keys are used to calculate secretly shared cover value.
Although symmetric keys are usually used for a longer time intervals, key renewal is a good practice to thwart cryptanalysis attacks.

As shown in Fig. \ref{fig:KeyAgreement}, to share a long-term secret symmetric key with the Utility company, each smart meter $SM_i$ chooses a random element $k_{i,u}\in Z^*_q$ and forms a key establishment request packet.
As shown in Fig. \ref{fig:KeyAgreement_1}, the packet has $ID_i, k_{i,u}P, TS_i$ and $\sigma_i$, where, $ID_i$ is the identifier of $SM_i$, $TS_i$ is the current timestamp, $k_{i,u}P$ is the random element ($k_{i,u}$ multiplied by the group generator $P$ of the additive group $G_1$), and $\sigma_i$ is signature of $SM_i$.
The smart meter then sends the key generation request packet to the Utility company.

The Utility company checks that the packet is not old by examining the timestamp $TS_i$ to prevent replay attacks.
Then, the Utility company checks the signature $\sigma_i$ on the information it has received to prevent man-in-the-middle attacks.
Then, the Utility company chooses a random element $k_{u,i} \in Z^*_q$ and calculates $k_{u,i}P$.
The Utility company calculates the long-term shared seed key $K_{u,i}$ to be equal to ($k_{u,i}k_{i,u}P$).
Finally, the Utility company sends the key establishment reply packet to $SM_i$.
As shown in Fig. \ref{fig:KeyAgreement_1}, the packet has $ID_u, k_{u,i}P, TS_u, H_2(K_{u,i}||1)$, and $\sigma_u$, where, $ID_u$ is the identifier of the Utility company, $TS_u$ is the current timestamp, $k_{u,i}P$ is the random element ($k_{u,i}$ multiplied by the group generator $P$), and $\sigma_u$ is the signature on the packet.
$H_2(K_{i,u}||1)$ is used for key confirmation.
When $SM_i$ receives the packet, it checks the timestamp and the signature similar to that of the Utility company.
Then, the smart meter $SM_i$ computes the long-term shared key $K_{u,i}$ to be equal to ($k_{i,u}k_{u,i}P$).
The derived keys by $SM_i$ and the Utility company are identical
\begin{equation*}
 K_{i,u} = K_{u,i} = k_{i,u}k_{u,i}P = k_{u,i}k_{i,u}P
\end{equation*}
Next, the smart meter $SM_i$ and the Utility company derive the session keys for each communication session.

%%% ----------------------- Figure ----------------------- %%%
\begin{figure}[t]
\setlength{\abovecaptionskip}{0mm}   % 0.5cm as an example
\setlength{\belowcaptionskip}{0mm}   % 0.5cm as an example
\centering
\includegraphics[scale=0.21]{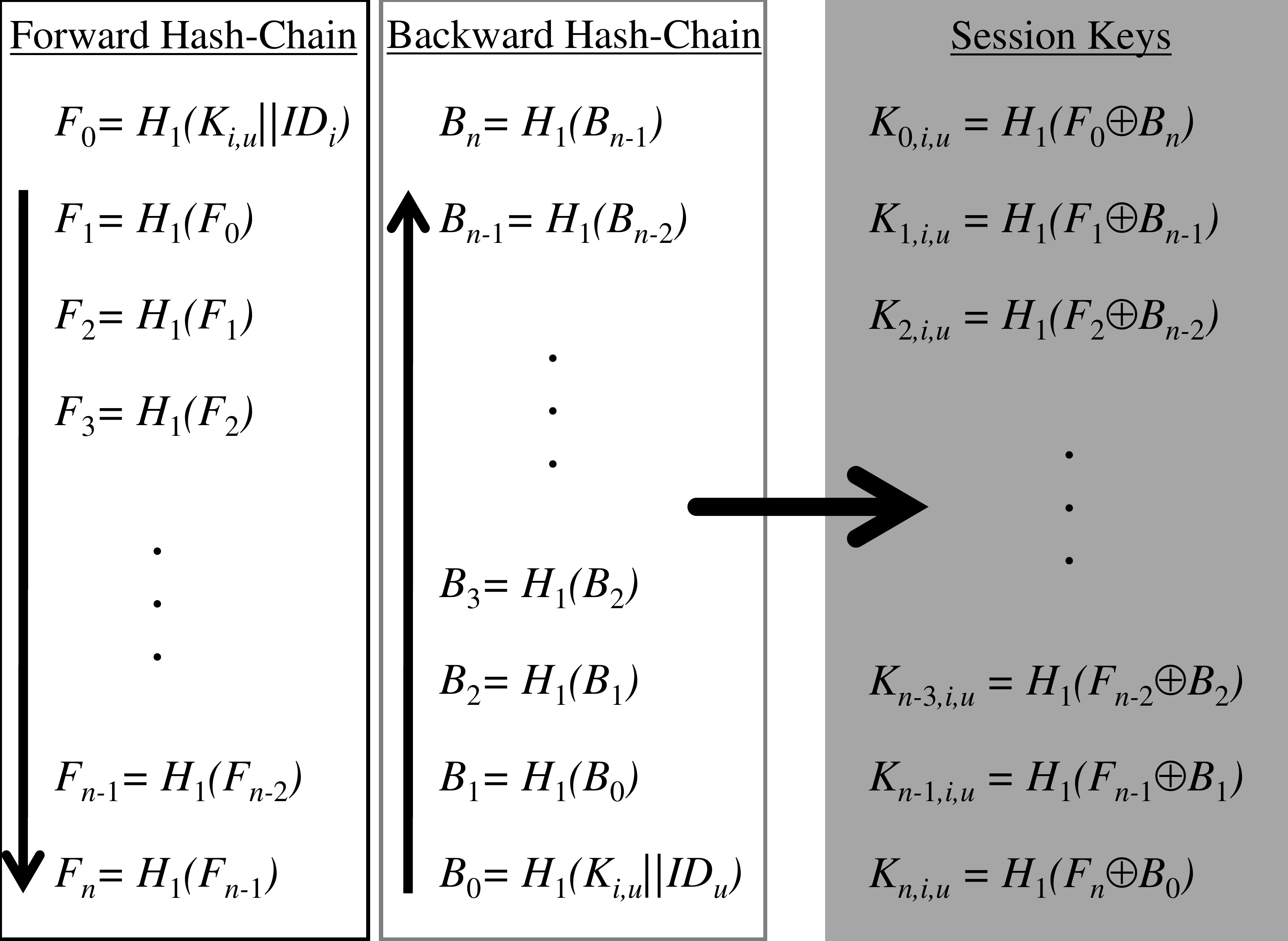}
\caption{Key Session generation steps at each elemnets in the AMI network.}
\vspace{-5mm}
\label{fig:KeyAgreement_2}
\end{figure}
%%% =========== Figure =========== %%%

The $K_{i,u}$ is the seed for the session keys.
The smart meter $SM_i$ and the Utility company use the seed to calculate a forward hash chain by using hash function $H_1$ on the seed key $K_{i,u}$.
As shown in Fig. \ref{fig:KeyAgreement_2}, the smart meter $SM_i$ and Utility company calculate $F_0$ by hashing $H_1(K_{i,u}||ID_i)$, then calculate $F_1$ by hashing $H_1(F_0)$ for $t$ time.
The last $F_t$ is equal to $H_1(F_{t-1})$.
To improve security, each key is only used for one communication session, and thus $t$ gives the number of the communication session messages that will be used in the long-term interval.
Then, the smart meter $SM_i$ calculates a backward hash chain for the same key $K_{i,u}$ just like above.
As shown in Fig. \ref{fig:KeyAgreement_2}, the smart meter $SM_i$ and the Utility company calculate $B_0$ by hashing $H_1(K_{i,u}||ID_u)$, then calculate $B_1$ by hashing $H_1(B_0)$ for $t$ time.
The last $B_t$ will be equal to $H_1(B_{t-1})$.
Finally, the smart meter $SM_i$ and the Utility company use
the hash function $H_1$ and XOR operation to calculate the session keys.
As illustrated in Fig. \ref{fig:KeyAgreement_2}, every key has two parts; one comes from the first forward hash chain and the other comes from the second backward hash chain.
For example, $K_{0,i,u}$ has two parts $F_0$ and $B_t$ that are XORed.
The first part comes from the beginning of the first hash chain $F_0$ and the second part comes from the end of the second hash chain $B_t$.
The $SM_i$ and the Utility company use the session keys to generate the same covered values.
After using all keys, they calculate a new long-term key $K_{i,u}$ and derive new session keys.

%%% ----------------------- Figure ----------------------- %%%
\begin{figure}[t]
\setlength{\abovecaptionskip}{-1mm}   % 0.5cm as an example
\setlength{\belowcaptionskip}{0mm}   % 0.5cm as an example
\centering
\includegraphics[scale=0.35]{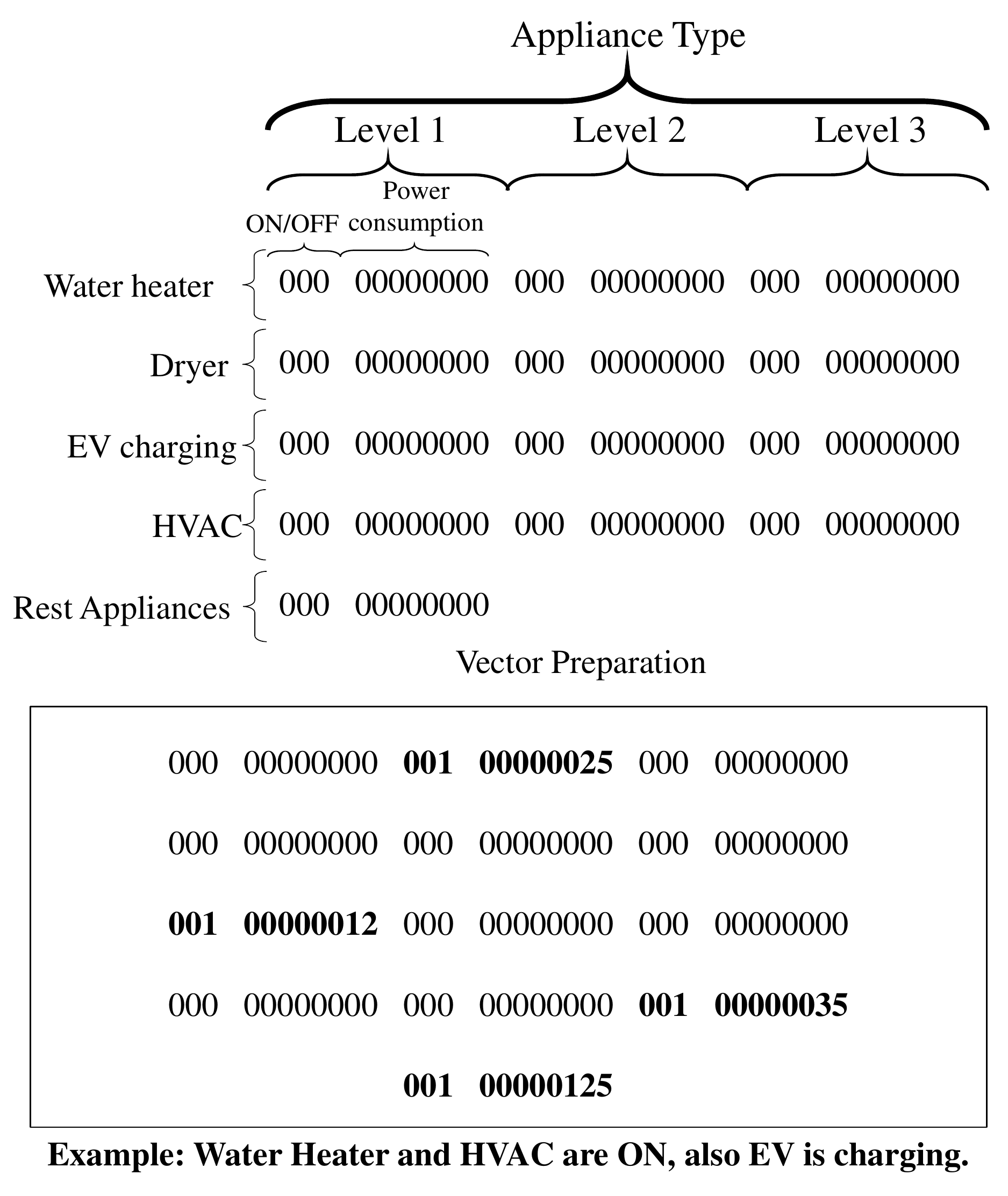}
\caption{Power consumption vector generation in our scheme.}
\vspace{-5mm}
\label{fig:Vector}
\end{figure}
%%% =========== Figure =========== %%%

\vspace{0mm}
\subsection{Power Consumption Vector Generation}
\label{Power Consumption Vector Generation}
\vspace{0mm}

Every smart meter $SM_i$ should make its power consumption vector that has the measure of energy consumed by each appliance.
The organization of the vector is given in Fig. \ref{fig:Vector}.
We divide the appliances in the home into two type, controllable and uncontrollable.
The controllable device such as (Water heater, Dryer, Electrical car charging, and Heating, ventilation, and air conditioning (HVAC)) can be controlled by the Utility company for total load managing.
The uncontrollable appliances are the rest of devices inside the home which the utility company cannot control it.
The vector divided into five groups, each group represent an appliance in the home.
Each appliance represents by three group of bits.
The first group for low power consumption, the second for medium power consumption and finally the third for high power consumption.
These three groups are divided into two subgroups.
The first part of the group of bits is for the number of appliances ON in the community, the second is for the amount of power consumed by these appliances.

When a controllable appliance is ON, the smart meter first checks the level of the appliance that it is working in, is it in low, medium or high level.
Then the smart meter according to the level of appliance's operation puts one in the number field and the amount of power consumption in the consuming field, as shown in Fig. \ref{fig:Vector}.
The smart meter does the above operation for all appliances.
In case the appliance is OFF, the smart meter puts zero in all fields of that appliance.
For uncontrollable appliances, the smart meter put one in the number field for the last group of bits then aggregate all the appliances reading and put the aggregated reading in the consumption field.

An example is shown in Fig. \ref{fig:Vector}, the bottom part of the figure illustrate that some appliances are ON inside the home and some of them are OFF.
The Water heater is ON, it is consuming 25 power unit and in the medium level power consumption.
Moreover, the Dryer is OFF, the Electrical Vehicle is charging, it is consuming 12 power unit in low power consumption level.
The HVAC is working in the high power consumption level and it is consuming 35 power units.
The rest of the house's appliances are consuming 125 power unit.

\vspace{0mm}
\subsection{Aggregating the power consumption reading and recovering the total reading from the covered messages}
\label{Aggregating the power consumption reading and recovering the total reading from the covered messages}
\vspace{0mm}

\begin{figure}[t]
\setlength{\abovecaptionskip}{0mm}   % 0.5cm as an example
\setlength{\belowcaptionskip}{0mm}   % 0.5cm as an example
\centering
\includegraphics[scale=0.29]{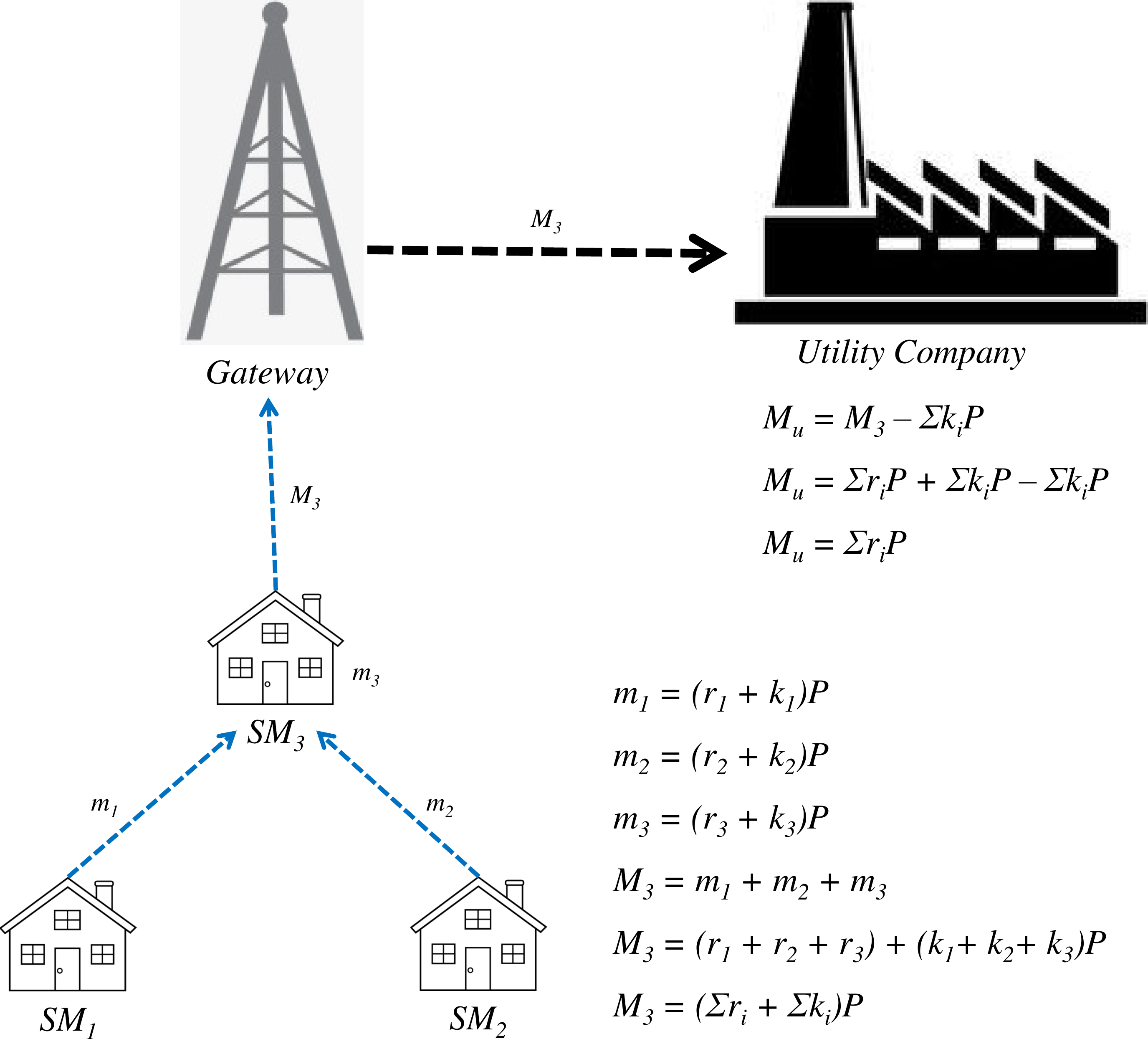}
\caption{Aggregate Reading from SM to Utility company and Utility company retrieve the total reading ($M_u$).}
\vspace{-5mm}
\label{fig:AggregateReading}
\end{figure}

Fig. \ref{fig:AggregateReading} describes a case on how the Utility company recovers the accumulated SMs' power consumption without knowing the individual power consumption readings.
For clearness purpose, we presume that the AMI system has just three smart meters, however, the concept can be stretched to more smart meters in the community.
We presume that the smart meter $SM_i$ has already a shared secret symmetric key with the Utility company.
As shown in the Fig. \ref{fig:AggregateReading}, each smart meter's message has a covered value that is the shared secret symmetric key with the Utility company $K_{i,u}$ that has been added to the power consumption reading $r_i$ vector and the result has been multiplied by the generator of the additive group $P$.
For instance, $SM_1$ send its power consumption covered reading $m_1$ to its parent smart meter $SM_3$.
The power consumption covered reading consist of the actual reading and the shared symmetric key with the Utility company ($m_1 = (r_1 + k_1)P$).
After $SM_3$ gets the messages of its children smart meters ($SM_1$ and $SM_2$), it adds the messages $m_1$ and $m_2$ to its own power consumption information $m_3$.

Given these two messages ($m_1$ and $m_2$), it is not feasible to recover the readings of $SM_1$ and $SM_2$ because $SM_3$ does not know the shared secret symmetric key with the Utility company $K_{i,u}$ that are added by smart meters.
$SM_3$ aggregates the messages of $SM_1$ and $SM_2$ with
its own covered reading to calculate $M_3$, where $M_3 = m_1 + m_2 + m_3 = (\Sigma r_i + \Sigma k_i)P$.
The scheme utilizes in-network accumulation, where the message sent by each smart meter has the power consumption of all its children smart meters.
Finally, the $SM_3$ sends $M_3$ to the Utility company that removes the covers to obtain the aggregated reading $\Sigma^3_{i=1}r_i$.

So, in general the Utility company removes the covers by subtracting all the shared secret symmetric key with all the smart meters in one batch operation.
The Utility company adds all the shared secret symmetric key to calculate $\Sigma^n_{i=1}K_{0,i,u}P$, then the Utility company subtracts this value from the accumulated power consumption that has been received from the Gateway
\begin{equation*}
M_u - \Sigma^n_{i=1}K_{0,i,u} = - \Sigma^n_{i=1}K_{0,i,u} + \Sigma^n_{i=1}K_{0,i,u} + \Sigma^n_{i=1}r_iP
\end{equation*}

The Utility company can get the accumulated power consumption information from each community and the total is $\Sigma^n_{i=1}r_i$, where $n$ is the number of the smart meters in that community.
All the messages should have hash values and timestamps to ensure message integrity and freshness.

\vspace{0mm}
\subsection{Load Control by Utility company}
\label{Load Control by Utility company}
\vspace{0mm}

After the Utility company gets the accumulated power consumption information from the community which is $\Sigma^n_{i=1}r_i$.
Then the Utility company checks the $\Sigma^n_{i=1}r_i$ vector to get an understanding of the numbers of the appliances that is power ON in the community and the level of their work.
In case the Utility company needs power reduction in any community due to overload or the Utility company has maintenance on one or more of its power generators, the Utility company replay the community through the gateway with a message and ask for a reduction in the controllable appliances.
The Utility company asks the smart meters that have the appliances that are working at the high-level to reduce power consumption more than the appliances that are at the medium-level.
If the power reduction in the medium and high-level is sufficient for the Utility company, then the Utility company does not ask low-level power consumption to reduce any power.
By dividing the appliances into controllable and uncontrollable, the Utility company can easily balance the load between generation and demand.
Also, by dividing each appliance into three level of operation we can ensure a fair reduction in power, the Utility company asks the high-level appliances to reduce more than the appliances that are in the medium-level and may not ask the appliances in the low-level to decrease any of their power consumption.
%%% =========== PROPOSED SCHEME =========== %%%

%%% ----------------------- Figure ----------------------- %%%
%%% =========== Figure =========== %%%

%%% ----------------------- Figure ----------------------- %%%
\begin{figure*}[ht!]
\setlength{\abovecaptionskip}{0mm}   % 0.5cm as an example
\setlength{\belowcaptionskip}{-5mm}   % 0.5cm as an example
\centering
\subfloat[Point Addition time completion for various number of SMs.\label{fig:ECC}]
  {\includegraphics[width=.33\linewidth]{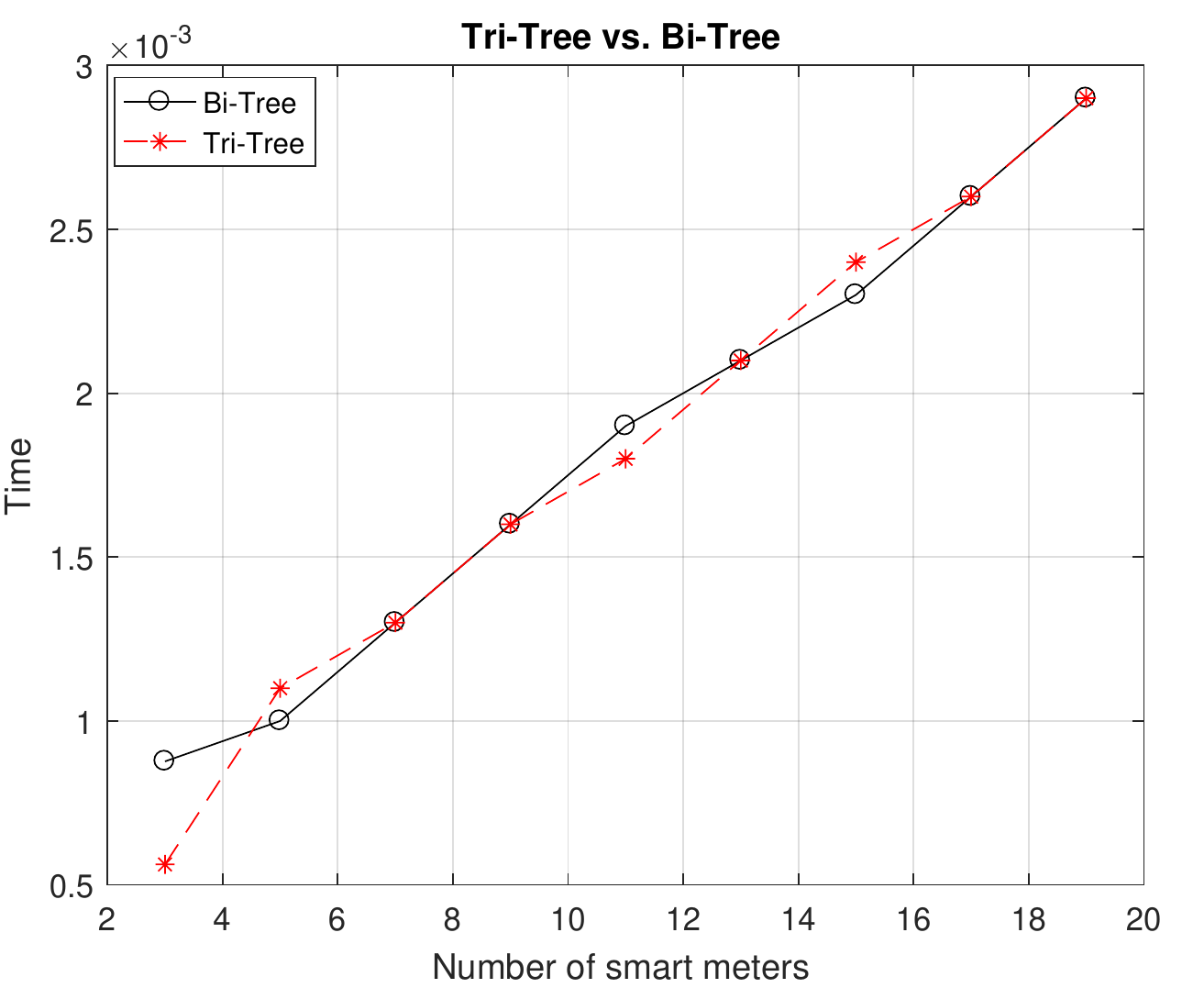}}\hfill
\subfloat[Homomorphic time completion  for various number of SMs.\label{fig:HE}]
  {\includegraphics[width=.33\linewidth]{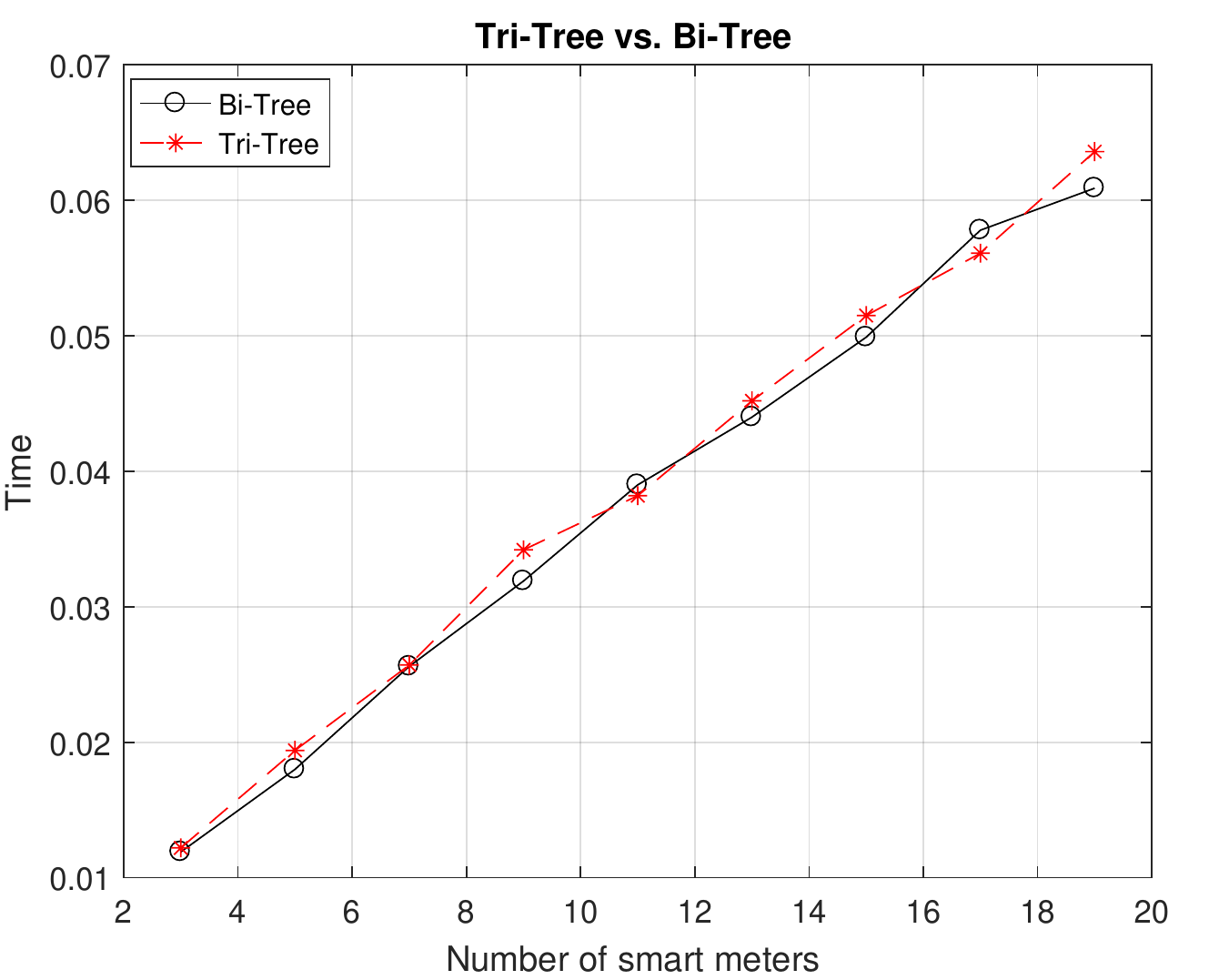}}\hfill
\subfloat[Point Addition vs Homomorphic time completion  for various number of SMs.\label{fig:HEvsECC}]
  {\includegraphics[width=.33\linewidth]{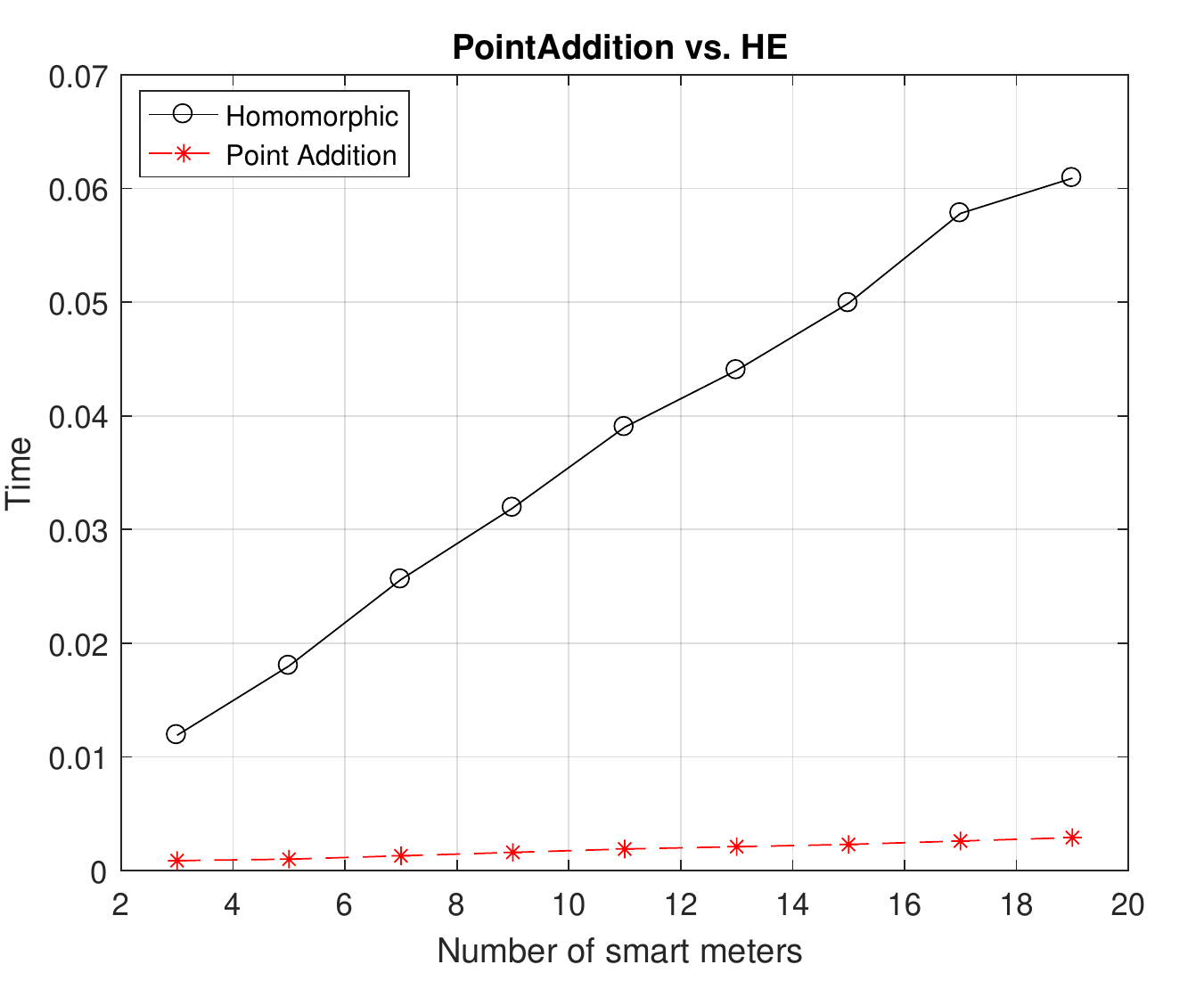}}
\caption{Performance comparison among Homomorphic and Point Addition}
\end{figure*}
%%% =========== Figure =========== %%%

%\begin{figure*}[htbp]
%\settoheight{\tempdim}{\includegraphics[width=0.3\textwidth]{Figs/ECC.pdf}}%
%%\rotatebox{90}{\makebox[\tempdim]{R=1}}\hfil
%\includegraphics[width=0.33\textwidth]{Figs/ECC.pdf}\hfil
%\includegraphics[width=0.33\textwidth]{Figs/ECC.pdf}\hfil
%\includegraphics[width=0.33\textwidth]{Figs/ECC.pdf}
%
%\medskip
%\hspace{0.65\baselineskip}\hfil
%\makebox[0.3\textwidth]{(a) \label{fig:ECC}}\hfil
%\makebox[0.3\textwidth]{(b) \label{fig:HE}}\hfil
%\makebox[0.3\textwidth]{(c) \label{fig:HEvsECC}}
%
%\caption{(a) Point Addition time completion for various number of SMs, (b) Homomorphic time completion  for various number of SMs, (c) Point Addition vs Homomorphic time completion  for various number of SMs.}%\label{fig}
%\end{figure*}

%%%% ----------------------- RESULTS ----------------------- %%%
\vspace{0mm}
\section{Results}
\label{Results}
\vspace{0mm}

In this paper, we address the privacy issue in the AMI network.
We divide home appliances into two categories, controllable and unconrtollable.
The controllable that can be controlled by the Utility company and the Utility company can know the number of specific appliances is ON in a community, the amount of power consumption by these appliances and the range of their working.

We choose point addition and homomorphic for adding the vectors of the consumer ($SM_i$).
As shown in Fig. \ref{fig:HEvsECC}, the homomorphic encryption need more time than that needed by point addition for the same network and topology.
As the homomorphic messages are longer and time for encryption and decryption need by homomorphic encryption is much more compared to the point addition, homomorphic encryption time for completion increase linearly with the increase of the number of the smart meters in the AMI network.
For the point addition, the trend is almost constant and the increment of the number of the smart meters in the AMI network will not affect the time to completion.

There is a trade-off between increasing the number of the smart meters in the AMI network and the overhead in the computation.
When the network is very big the time to completion will increase with the homomorphic encryption, as shown in Fig. \ref{fig:HEvsECC}, but time for decryption stay the same as we need one decryption for the accumulated message.
on the other hand, increasing the number of the smart meters in the AMI network using addition point does not increase in the completion time but retrieving the message from the accumulated message ($\Sigma^n_{i=1}r_i$P) will increase.
So, for a small network, the point addition gives a good result as need less time for converging and less time for retrieving the data.
but for a big network, the homomorphic encryption gives a better result.
The homomorphic encryption needs more time for converging but retrieving the data is a one decryption operation.

We test our scheme (point addition and homomorphic encryption) on a tree topology as it is most common topology.
We choose a binary tree that each smart meter has two children and also on Tri-tree where the smart meter has three children.
As shown in Figs. \ref{fig:ECC} and \ref{fig:HE}, both operation (point addtion and homomorphic encyption) has almost the same performance on the two topology.
But the performance change with changing the encryption method (point addition and homomorphic) as shown in Fig. \ref{fig:HEvsECC}.

After the Utility company gets the accumulated message, the Utility company decrypt the message and from the result, the Utility company knows the number of appliances (controllable) that are ON in the community and the Utility company know the power consumption per appliance.
This result helps the Utility company to ask for the reduction in power consumption without knowing the individual reading.

%%%% =========== RESULTS =========== %%%

%%%% ----------------------- RELATED WORK ----------------------- %%%
%\vspace{0mm}
%\section{Related Work}
%\label{Related Work}
%\vspace{0mm}
%
%Authors in [1] provided an analytical framework to evaluate the impact of plug-in hybrid electric vehicle (PHEV) loading on a distribution system and used stochastic methods to study the possible impacts of charging EVs on distribution network components.
%Authors in [8] considered case studies with different EV penetration levels and charging patterns and estimated the maximum number of EVs that a distribution network can accommodate based on an N-1 contingency condition.
%Authors in [9] and [10] focused on the distribution system losses due to EV charging.
%In [9] an optimal EV fleet charging profile is proposed for minimizing the distribution system power losses.
%For even smaller sized networks, authors in [11] and [12] emphasized on the integration of EVs at the distribution transformer level serving a few houses and proposed household load control strategies to tackle the transformer overloading problem.
%%%% =========== RELATED WORK =========== %%%

%%% ----------------------- CONCLUSION ----------------------- %%%
\vspace{0mm}
\section{Conclusion}
\label{Conclusion}
\vspace{0mm}

We have proposed an effective scheme to empower the Utility company to gather the power consumption information while saving the consumers' privacy.
Our scheme generally utilizes lightweight symmetric key cryptography and hashing operations to gather power consumption information.
The asymmetric key cryptography operation is required just for the key management that is executed each lengthy time-frame.
Our estimations have demonstrated that our scheme can decrease the package measure, as well as lessen the calculation time at each smart meter.
We utilized MatLab system to evaluate the effect of this overhead decrease on the system execution.
The simulation has exhibited that our scheme expends considerably less transmission capacity, has less package transmission, and less delay.
Our investigation has shown that the proposed scheme is secure and can save consumers' privacy.
%%% =========== CONCLUSION =========== %%%
%

\vspace{0mm}

\bibliographystyle{IEEEtran}
\bibliography{DR}
\end{document}